\documentclass[10pt,english, aps,superscriptaddress,showpacs,floatfix,prl,lengthcheck,floatfix,nofootinbib,twocolumn]{revtex4-1}

\usepackage[T1]{fontenc}
\usepackage[latin9]{inputenc}
\usepackage{color}
\usepackage{textcomp}
\usepackage{amsmath}
\usepackage{amssymb}
\usepackage{graphicx}

\makeatletter
\usepackage[latin9]{inputenc} %%% acentos no bibtex resolve a questao

\usepackage{braket}

\usepackage{amsmath}

\usepackage[T1]{fontenc}
\usepackage{color}
\usepackage{babel}
\usepackage{verbatim}
\usepackage{amsmath}
\usepackage{amssymb}
\usepackage{graphicx}
\usepackage[pdfusetitle,colorlinks,linkcolor=blue,citecolor=blue,urlcolor=blue]{hyperref}
\usepackage{microtype}
\usepackage{braket}

\makeatother

\usepackage{babel}
\begin{document}
\title{Light-wave coherent control of the insulator-to-metal transition in
a strongly correlated material}
\author{Eduardo B. Molinero}
\email{ebmolinero@gmail.com}

\affiliation{\emph{Instituto de Ciencia de Materiales de Madrid, Consejo Superior
de Investigaciones Científicas (ICMM-CSIC), Madrid, Spain}}
\author{Rui E. F. Silva}
\email{rui.silva@csic.es, ruiefdasilva@gmail.com}

\affiliation{\emph{Instituto de Ciencia de Materiales de Madrid, Consejo Superior
de Investigaciones Científicas (ICMM-CSIC), Madrid, Spain}}
\affiliation{\emph{Max-Born-Institut, Max Born Strasse 2A, D-12489 Berlin, Germany}}
\begin{abstract}
The use of intense tailored light fields is the perfect tool to achieve
ultrafast control of electronic properties in quantum materials. Among
them, Mott insulators are materials in which strong electron-electron
interactions drive the material into an insulating phase. When shinning
a Mott insulator with a strong laser pulse, the electric field may
induce the creation of doublon-hole pairs, triggering an insulator-to-metal
phase transition. In this work, we take advantage of the threshold
character of this insulator-to-metal transition and we propose a pump-probe
scheme that consists of a mid-infrared laser pulse and a train of
short pulses separated by half-period of the mid-infrared with alternating
phases. By varying the time-delay between the two pulses and the internal
carrier envelope phase of the short pulses, we achieve control of
the phase transition,\textcolor{black}{{} which leaves its fingerprint
at its high harmonic spectrum.}
\end{abstract}
\maketitle
The latest developments in laser science have unlocked the possibility
to engineer intense ultrashort pulses that are able to steer and control
the dynamics of electrons, both in atoms and molecules \citep{krausz2009attosecond}
as well as in solid state materials \citep{kruchinin2018colloquium}.
When the light-matter interaction is comparable with the Coulomb potential
created by the nuclei, highly non-linear optical phenomena start to
appear. In particular, strong field ionization and high harmonic generation
are the two fundamental processes in strong field physics \citep{keldysh1965ionization,corkum1993plasma,lewenstein1994theory,ivanov2005anatomy}
and have played a key role in our understanding of ultrafast electron
dynamics \citep{ivanov2005anatomy,smirnova2009high,lepine2014attosecond}. 

The rapid development of lightwave engineering allowed the emergence
of several laser techniques that are able to probe ultrafast dynamics
at the attosecond timescale, such as streaking \citep{constant1997methods,hentschel2001attosecond,itatani2002attosecond,kaldun2016observing}
and RABBIT (reconstruction of attosecond beating by interference of
two-photon transitions) \citep{paul2001observation,silva2012autoionizing,riviere2012pump,gruson2016attosecond}.
Both techniques use a pump-probe setup, an attosecond pulse train
(RABBIT) or a single attosecond pulse (streaking) as pump and an infrared
pulse (attosecond pulse train) as a probe. 

Recently, there has been an increasing effort to port the attosecond
techniques that were developed in the context of atoms and molecules
to phenomena that are relevant in the condensed phase \citep{kruchinin2018colloquium,ghimire2019high}.
The discovery of high harmonic generation in ZnO \citep{ghimire2011observation}
triggered the use of high harmonic spectroscopy in the condensed phase
allowing for: the all-optical reconstruction of bands \citep{vampa2015all},
the observation of Bloch oscillations \citep{schubert2014sub,ghimire2011observation},
electron-hole dynamics \citep{vampa2014theoretical,hohenleutner2015real,yue2020imperfect},
van-Hove singularities \citep{uzan2020attosecond}, topological phase
transitions \citep{silva2019topological,chacon2020circular,baykusheva2021all,bai2021high,schmid2021tunable}
and dynamics in strongly correlated materials \citep{Silva2018,bionta2021tracking,tancogne2018ultrafast,murakami2018high,imai2020high,murakami2021high,shao2022high,orthodoxou2021high,zhu2021ultrafast}. 

\begin{figure}
\begin{centering}
\includegraphics[width=0.8\columnwidth]{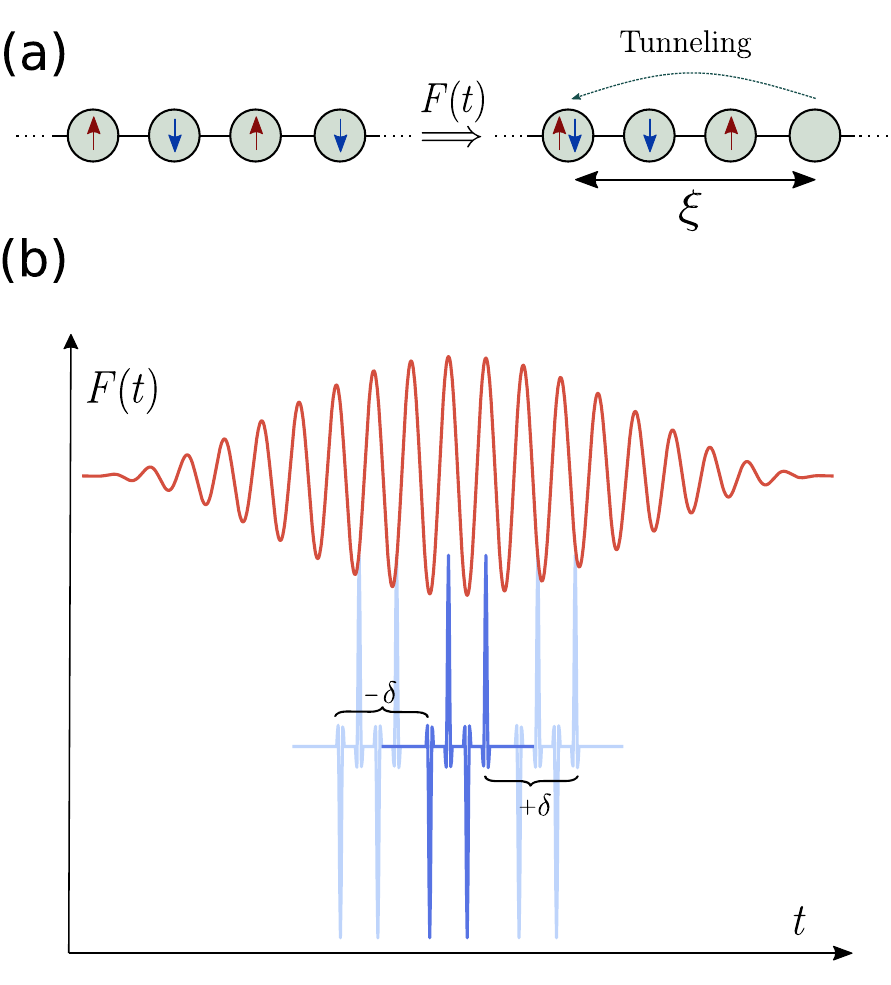}
\par\end{centering}
\caption{\label{fig:1}(a) Schematic representation of the tunneling phenomenon
in the 1D Fermi-Hubbard model. (b) Representation of the two different
lasers. The red line denotes the mid-IR laser while the blue line
corresponds to the train of short pulses with $\phi_{\text{CEP}}=0$.
Lighter blue lines shows how a $\pm\delta$ delay affects the train.}
\end{figure}

\begin{figure*}
\includegraphics[width=2\columnwidth]{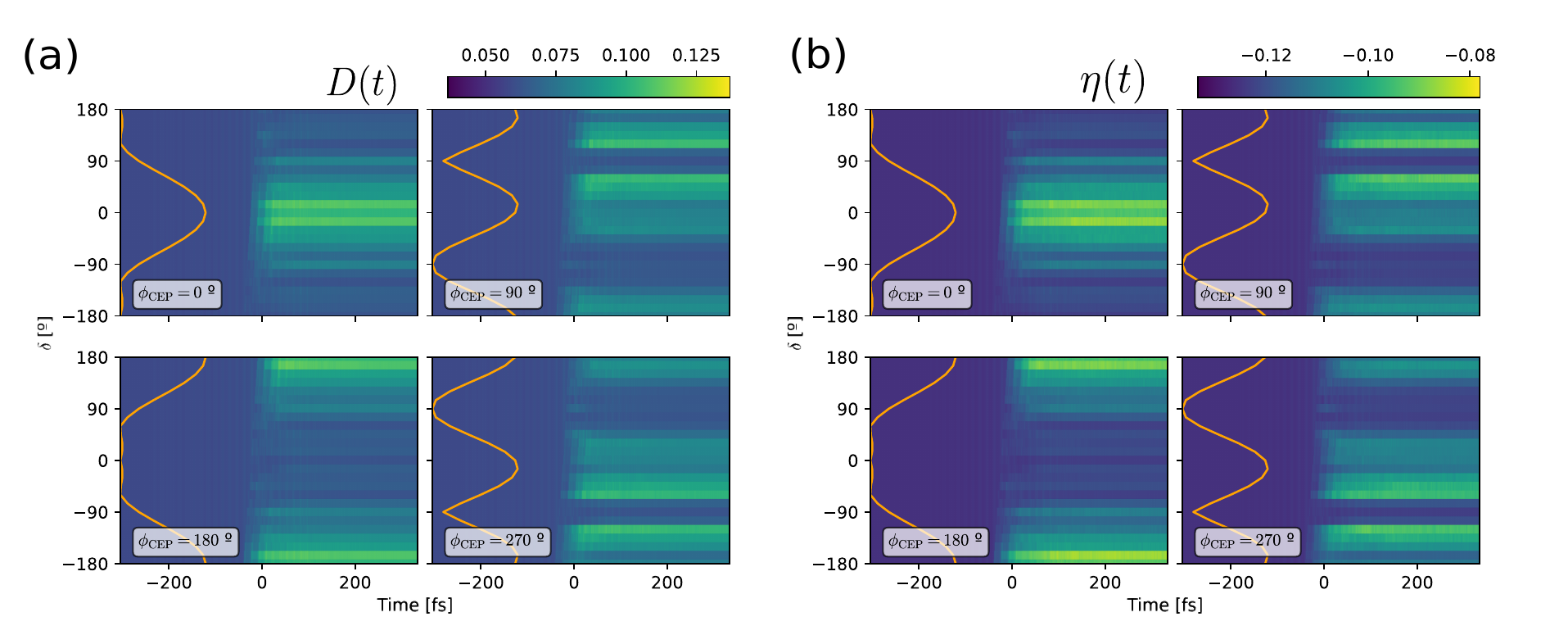}

\caption{\label{fig:2_time_evo}Average number of doublon-hole pairs, $D\left(t\right)$,
(a) and next-neighbour spin--spin correlation function, $\eta\left(t\right)$,
(b) as a function of time and $\delta$, for different values of $\phi_{\mathrm{CEP}}$.
The values shown are convoluted with a normal distribution function
with $\sigma=1$ fs. The orange line shows the total production rate
$\Gamma$ obtained for the corresponding delay. The CEP of the laser
is shown inside each figure.}
\end{figure*}

In a strongly correlated material the formation of doublon-hole pairs
under the presence of a strong electric field has a threshold character
\citep{oka2012nonlinear}. Previous works \citep{mayer2015tunneling,yamakawa2017mott,mazza2016field}
have shown that we can induce a dynamical phase transition on the
system by varying the field strength $F_{0}$ of the electric field
(given a fixed interaction parameter $U$). More precisely, the laser
can break the antiferromagnetic order of the Mott insulator by favoring
the appearance of doublon-hole pairs. The mechanism behind this phenomenon
is analogous to that present in strong field ionization in atoms.
In fact, one could define an adiabaticity parameter $\gamma_{K}$
which plays the same role as the Keldysh parameter in atoms \citep{keldysh1965ionization,ivanov2005anatomy},
\textcolor{black}{this parameter is defined} as $\gamma_{K}=\hbar\omega_{L}/\xi F_{0}$
where $\xi$ is the correlation length \citep{oka2012nonlinear} and
$\omega_{L}$ is the frequency of the laser. In the \emph{tunneling}
regime, $\gamma_{K}\ll1$, an electron can tunnel through the interaction
repulsion over an distance $\sim\xi$ due to the presence of the laser.
\textcolor{black}{Thus, a doublon-hole pair is formed, which leads
to the melting of the insulator state}, see Fig. \ref{fig:1}(a).
However and due to its threshold character, this will only happens
if the field goes higher than a certain value $F_{\text{th}}$ \citep{oka2012nonlinear,Silva2018}.
In Mott-like systems, this threshold is given by $F_{\text{th}}=\Delta/2e\xi$
where $\Delta$ is the Mott gap and $\xi$ is the correlation length
\citep{oka2012nonlinear}.

In this Letter, we propose the use of a pump-probe scheme, similar
to the one used in RABBIT, that consists of a femtosecond pulse train
and a mid-infrared pulse, see Fig. \ref{fig:1}(b), and by taking
advantage of the threshold character of the insulator-to-metal transition
to be able to control the insulator-to-metal transition in a strongly
correlated material. 

\begin{figure*}
\includegraphics[width=2\columnwidth]{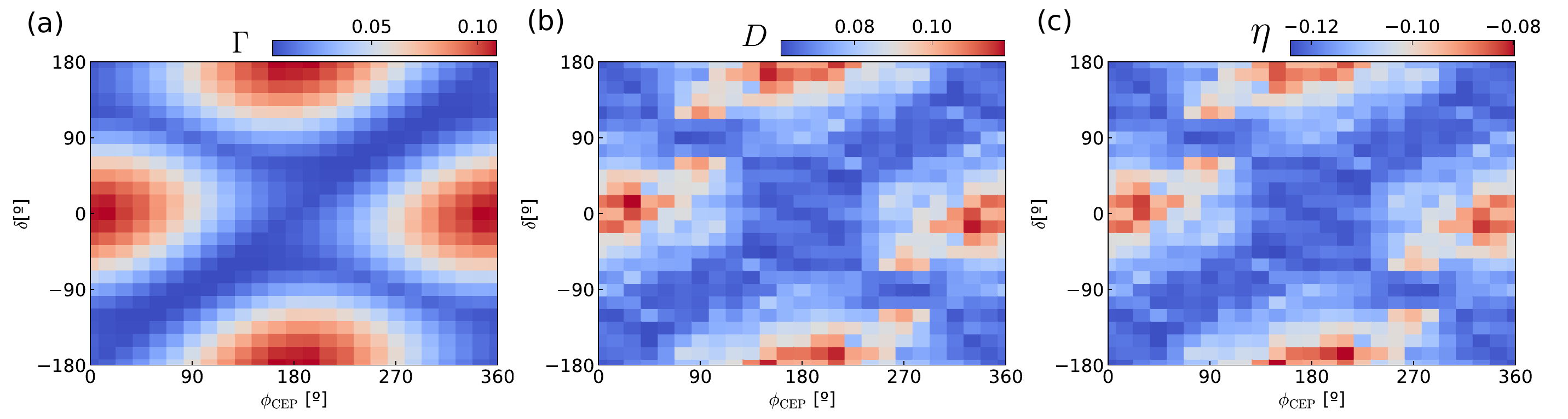}

\caption{\textbf{\label{fig:3_phase_diagram}}(a): Total production rate $\Gamma$.
Phase diagrams of the $D\left(t\right)$ (b) and $\eta\left(t\right)$
(c). Both these phase diagrams are obtained as the mean value on a
30 fs period of time after the end of the laser pulse.}
\end{figure*}

\emph{Setup}.---We will focus on the one-dimensional Hubbard model
and we will use parameters to mimic $\mathrm{Sr_{2}CuO_{3}}$. The
main purpose of the present work is to show that one can take advantage
of this tunneling phenomenon, in order to acquire a coherent control
over the insulator-to-metal transition in a strongly correlated material. 

In order to model the dynamics, we consider the one dimensional Hubbard
model, 
\begin{align}
H= & -\tau\sum_{j=1,\sigma}^{L}(e^{-\mathrm{i}\Phi(t)}c_{j,\sigma}^{\dagger}c_{j+1,\text{\ensuremath{\sigma}}}+\mathrm{h.c.})\nonumber \\
 & +U\sum_{j=1}^{L}n_{j,\uparrow}n_{j,\downarrow},
\end{align}
where $c_{j,\sigma}^{\dagger}(c_{j,\sigma})$ is the fermionic creation
(annihilation) operator for site $j$ and spin $\sigma$ and $n_{j,\sigma}=c_{j,\sigma}^{\dagger}c_{j,\sigma}$
is the number operator. The physical parameters $\tau$ and $U$ have
been set to $\tau=0.52$ eV and $U=5.96\tau$ to mimic $\mathrm{Sr_{2}CuO_{3}}$
\citep{oka2012nonlinear}, which also ensures us that we are in the
strong coupling limit, i.e., $U\gg\tau$. We will focus on the half-filling
case (one electron per site). In this case, the ground state (also
referred as the Mott insulator) has a short-range antiferromagnetic
order \citep{essler2005hubbard}; the electrons become localized in
position space with anti-parallel spins respective to their adjacent
sites. The laser electric field, $F(t)$, is taken into account through
the Peierls phase \citep{essler2005hubbard}: $\mathrm{d}\Phi(t)/\mathrm{d}t=-eaF(t)$,
where $a=7.56$ a.u. is the lattice constant of $\mathrm{Sr_{2}CuO_{3}}$
and $e$ is the electron charge. We will solve it using $N=L=12$,
setting periodic boundary conditions ($c_{j+L,\sigma}=c_{j,\sigma}$)
and focusing on the $S_{z}=0$ subspace. Our approach consist of exactly
solving the time-depedent Schrödinger equation (TDSE) using a timestep
of $dt=0.5$ a.u.. Convergence with the number of sites was checked
by performing calculations where $N=L=14$.

As previously mentioned, we will use two different lasers (in a pump-probe
scheme) and\textcolor{red}{{} }tune the time delay $\delta$ between
them in order to obtain the control. The first of them will consist
on a mid-IR laser with frequency $\omega_{\text{IR}}=32.92$ THz and
an amplitude of $F_{0,\text{IR}}=5$ MV/cm. The second laser will
be a train made up of 4 short pulses (their total duration is roughly
$6$ fs) equally splitted. The time-delay between 2 consecutive short
pulses is half-period of the mid-IR laser pulse and they have alternating
phases, in accordance to what is obtained by means of high harmonic
generation. The parameters for the short pulses are $\omega_{\text{pul}}=5\omega_{\text{IR}}=164.6$
THz, having a central frequency that corresponds to the 5$^{\mathrm{th}}$
harmonic, and a amplitude of $F_{0,\text{pul}}=8$ MV/cm. Both lasers
are modulated by a $\cos^{2}$ envelope \citep{joachain2009atoms}
as it can be appreciated in Fig. \ref{fig:1}(b). If one defines a
single laser pulse as 
\begin{align}
E\left(t,t_{0},T,\omega,E_{0},\phi\right)= & E_{0}\cos^{2}\left(\pi\left(t-t_{0}\right)T^{-1}\right)\nonumber \\
\theta\left(0.5T-\left|t-t_{0}\right|\right) & \cos\left(\omega\left(t-t_{0}\right)+\phi\right),
\end{align}
the pump-probe scheme used is given by 
\begin{align}
F\left(t\right) & =E\left(t,0,20T_{\text{IR}},\omega_{\text{IR}},F_{0,\text{IR}},\pi/2\right)\nonumber \\
+ & \sum_{n=1}^{4}E\left(t,t_{0,n},T_{\mathrm{pul}},\omega_{\text{pul}},F_{0,\text{pul}},\phi_{\mathrm{CEP}}-n\pi\right),
\end{align}
where $\omega_{\text{IR}}=32.92$ THz is the frequency of the mid-IR
pulse ($T_{\mathrm{IR}}=2\pi/\omega_{\mathrm{IR}}$), $\omega_{\text{pul}}=5\omega_{\text{IR}}$
is the central frequency of the single-cycle femtosecond pulses ($T_{\mathrm{pul}}=2\pi/\omega_{\mathrm{pul}}$),
$F_{0,\text{IR}}=5$ MV/cm is the amplitude of the mid-IR pulse, $F_{0,\text{pul}}=8$
MV/cm is the amplitude of the single-cycle femtosecond pulses, $t_{0,n}=\left(\left(2n-5\right)/4\right)T_{\mathrm{IR}}+\delta/\omega_{\mathrm{IR}}$,
$\delta$ is the delay between the two pulses in radians and $\phi_{\mathrm{CEP}}$
is the carrier envelope phase of the single-cycle femtosecond pulses.

The threshold field \citep{oka2012nonlinear} for $\mathrm{Sr_{2}CuO_{3}}$
takes a value of $F_{\text{th}}=9.1$ MV/cm so neither of the fields
can surpass it on their own; only when the two amplitudes sum in a
coherent way, namely when the short pulses land right on the peaks
of the mid-IR, the electric field will break the threshold and thus,
the phase transition will take place. The specific delays in which
this will occur will heavily depend on carrier-envelope phase (CEP),
denoted by $\phi_{\text{CEP}},$ of the short pulses. For example,
by looking at Fig. \ref{fig:1}(b) one can easily guess that for $\phi_{\text{CEP}}=0$,
the two laser will sum coherently at $\delta=0$. As a result, we
can \emph{control the phase transition} by light-wave engineer of
the two parameters $\delta$ and $\phi_{\text{CEP}}$. It must be
emphasized that these laser parameters are well within experimental
reach.

To characterize the insulator-to-metal transition, one must pay attention
to both charge and spin degrees of freedom. To do so, we will compute
the following observables in the Schrödinger picture. Firstly, the
average number of doublon-hole pairs

\begin{equation}
D(t)=\frac{1}{L}\sum_{j}\left\langle n_{j,\uparrow}n_{j,\downarrow}\right\rangle 
\end{equation}
and secondly, the next-neighbour spin--spin correlation function
\begin{equation}
\eta(t)=\frac{1}{L}\sum_{j}\left\langle \vec{S}_{j}\cdot\vec{S}_{j+1}\right\rangle ,
\end{equation}
where $\vec{S}_{j}$ is the vector of spin matrices for spin $1/2$
and site $j$. 

At first, it may seems that predicting for which values of $\delta$
and $\phi_{\text{CEP}}$ the phase transition takes place can be quite
a task. However, we can accomplish such thing by just computing the
maximum production rate $\Gamma$ of the doublon-hole pairs in \emph{the
tunneling regime}. Following Ref. \citep{oka2012nonlinear}, we take
the production rate as
\begin{equation}
\Gamma=\exp\text{\ensuremath{\left(-\ensuremath{\pi\frac{F_{\text{th}}}{\max|F(t)|}}\right)}}.\label{eq:gamma}
\end{equation}

Therefore, only by knowing the shape of the laser and the threshold
amplitude (which depends solely on the material), we can do a very
confident guess of the non-equilibrium behaviour of the material.
It must be noted that this rough estimation is only valid when we
are in the tunneling regime. By calculating the adiabaticity parameter
for both pulses separately, we found that $\gamma_{K,\mathrm{pul}}=1.03$
and $\gamma_{K,\mathrm{IR}}=0.33$. The mid-IR pulse is well in the
tunneling regime and the train pulse is in the frontier between multiphoton
and tunneling regime. However, when calculating $\gamma_{K,\mathrm{pul}}$
summing the two field strengths we get $\gamma_{K,\mathrm{pul}}=0.63$,
which is indeed in the tunneling regime. Consequently, $\Gamma$ as
defined in Eq. (\ref{eq:gamma}), for our laser parameters, can be
used as a good number indicating whether or not we are inducing an
insulator-to-metal transition.

\emph{Results}.---To prove the above assumptions, we have performed
several numerical calculations. In Fig. \ref{fig:2_time_evo}(a) we
can see that the profile of $D(t)$ coincides almost perfectly with
the prediction given by $\text{\ensuremath{\Gamma}}$. When the production
rate starts to increase, i.e., when the delay $\delta$ (and the $\phi_{\text{CEP}}$)
causes the field to fulfill $\max|F(t)|>F_{\text{th}}$, doublon-hole
pairs begin to appear due to the tunneling mechanism and so, the insulator
state breaks down, leaving the system in a photo-induced saturated
state \citep{oka2008photoinduced,oka2012nonlinear}. On the other
hand, when $\delta$ and $\phi_{\text{CEP}}$ are such so that the
field do not surpass the threshold, the initial ground state is kept
intact. The same trend can be appreciated for $\eta(t)$, the antiferromagnetic
order of the Mott state is lost when $\Gamma$ becomes maximum and
is conserved in the opposite case (Fig. \ref{fig:2_time_evo}(b)).
Despite the simplicity of Eq. (\ref{eq:gamma}), it does capture the
physics of the system in a remarkable manner, predicting even abrupt
changes in the phase of the system (Fig. \ref{fig:2_time_evo}(a)
and (b) corresponding to the case $\phi_{\text{CEP}}=90\text{º}$).
This resemblance between $\Gamma$ and the physical observables reaffirms
that the dynamics of the system are heavily dominated by the tunneling
mechanism.

The possibility of making accurate predictions along with the tunability
of the two parameters $\delta$ and $\phi_{\text{CEP}}$, is what
gives us the \emph{control over the Mott transition }as shown in Fig.
\ref{fig:3_phase_diagram}, where we have included $\Gamma$ alongside
the phase diagram of the system for a range of $\delta$ and $\phi_{\text{CEP}}$.
Comparing Fig. \ref{fig:3_phase_diagram}(a) with (b), one can see
the clear correspondence between the theoretical prediction of the
production rate and the obtained phase. As stated previously, when
the production rate reaches a maximum, the system undergoes a phase
transition if not, the system stays in the Mott insulator state. Nevertheless,
and in spite of the similarities between the two quantities, they
do not show exactly the same behavior. This is because tunneling is
not a deterministic process but rather a probabilistic one; the electron
is \emph{not guaranteed} to tunnel over the potential barrier. Also,
is worth noting the expected periodicity both in the delay and in
the CEP.

\begin{figure}
\includegraphics[width=1\columnwidth]{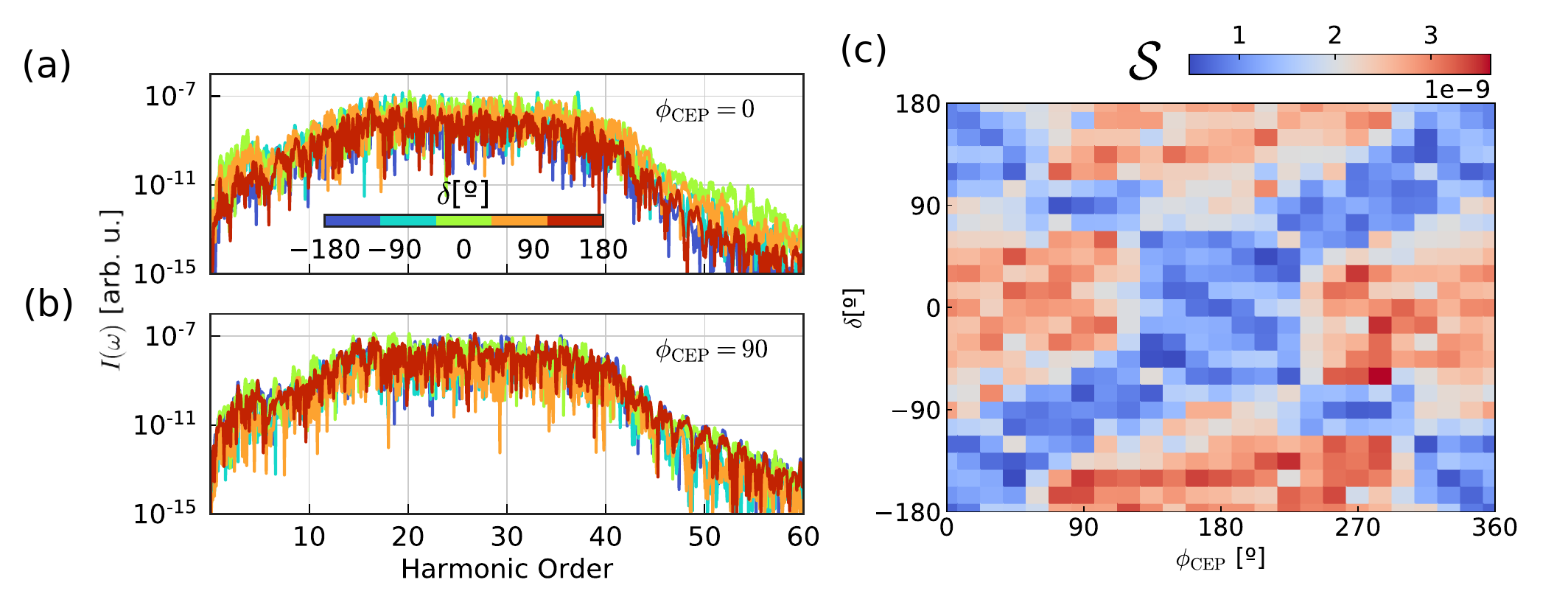}

\caption{\label{fig:4_hhg_energy}Harmonic spectra in arbitrary units for $\phi_{\text{CEP}}=0$
(a) and for $\phi_{\text{CEP}}=90$ (b). We have included only five
different delays in each plot for the sake of clarity. (c) Integrated
spectrum $\mathcal{S}$ computed for the same range of $\delta$ and
$\phi_{\text{CEP}}$ as in Fig. \ref{fig:3_phase_diagram}.}
\end{figure}

However, measuring correlations in Mott-like systems is not an easy
task \citep{dean2016magnetic_corr}. Therefore, if we want to engineer
the phase transition it is almost mandatory to find a more suitable
figure of merit. Naturally, and since we are generating optical charge
excitations, this process must manifest itself in the optical response
of the system. Indeed, previous works have shown that the insulator-to-metal
transition induced by strong laser fields leaves a fingerprint in
the corresponding harmonic emission \citep{Silva2018,orthodoxou2021high}.
To compute the optical emission, we first obtain the electric current
operator \citep{essler2005hubbard}
\begin{equation}
\hat{J}(t)=-\mathrm{i}ea\tau\sum_{j,\sigma}(e^{-\mathrm{i}\Phi(t)}c_{j,\sigma}^{\dagger}c_{j+1,\text{\ensuremath{\sigma}}}-\mathrm{h.c.}).
\end{equation}
Afterwards, the harmonic spectra is computed through the Larmor's
formula $I(\omega)\propto\omega{{}^2}|\braket{\hat{J}(\omega)}|^{2}$.
The spectrum (Fig. \ref{fig:4_hhg_energy}) displays the same features
presented in \citep{Silva2018}. The low intra-band harmonics are
mostly suppressed while the high harmonics are the most prominent.
Furthermore, these are centered around the harmonic $N=U/\omega_{L}\sim21$
as expected. However, if we compare the emission between different
CEPs (Figs. \ref{fig:4_hhg_energy} (a) and (b)) there is no significant
difference between $\phi_{\text{CEP}}=0\text{º}$ and $\phi_{\text{CEP}}=90\text{º}$.
Additionally, varying the delay doesn't give a different spectra either. 

However, if one now computes the integrated spectrum 
\begin{equation}
\mathcal{S}=\int_{\omega_{-}}^{\omega_{+}}\text{d}\omega\,I(\omega),\label{eq:inte_energy}
\end{equation}
a different behavior can be appreciated. We first note that the frequencies
$\omega_{+}$ and $\omega_{-}$ corresponds to the upper and lower
limits of the energy of the first allowed optical excitations \citep{oka2012nonlinear,essler2005hubbard},
namely $\omega_{-}=\Delta$ and $\omega_{+}=\Delta+8\tau$. In Fig.
\ref{fig:4_hhg_energy}(c) we have obtained $\mathcal{S}$ for the
whole range of CEPs and delays, where it can be seen the noteworthy
resemblance between that quantity and the prediction made in Fig.
\ref{fig:3_phase_diagram}(a) using $\Gamma$. Accordingly, we can
also characterize the phase of the material thanks to its optical
response.

\emph{Conclusion}.---Our work has shown that is possible to acquire
control of the Mott transition in one dimensional systems using electric
fields. More specifically, we have shown that by superposing two different
lasers, a mid-IR and a train of short pulses, the phase transition
can be engineered by tuning the internal parameters of the lasers,
the time-delay between them and the internal carrier envelope phase
of the short pulses ($\delta$ and $\phi_{\text{CEP}}$ in our case).
Alongside this tunability, we proved that the total doublon-hole production
rate, $\Gamma$, gives a simple, yet accurate, method of predicting
the transition. Lastly, we found the existence of a more appropriate
figure of merit to characterize the transition experimentally, by
looking at the nonlinear optical response of the system. This work
may pave the way to experimental efforts in which the insulator-to-metal
transition in strongly correlated systems can be achieved in a coherent
way using tailored laser pulses.
\begin{acknowledgments}
The authors acknowledge fruitful discussions with Álvaro Jiménez-Galán
and Misha Ivanov. E. B. M. and R. E. F. S. acknowledge support from
the fellowship LCF/BQ/PR21/11840008 from \textquotedblleft La Caixa\textquotedblright{}
Foundation (ID 100010434).
\end{acknowledgments}

\end{document}